\documentclass[10pt,a4paper,twoside]{aa}
\usepackage{graphicx}
\usepackage{latexsym}
\usepackage{amsmath}
\usepackage{amssymb}
\usepackage{amstext}
\usepackage{color}
\usepackage{psfrag}
\usepackage{natbib}
\def\m{$^\prime$}
\def\s{$^{\prime\prime}$~}
\def\hh{$^{\mathrm h}$}
\def\mm{$^{\mathrm m}$}
\def\ss{$^{\mathrm s}$}
\def\c2{cm$^{-2}$}
\def\cm3{cm$^{-3}$}
\def\pp{^{\prime\prime}}
\def\12{$^{12}$CO}

\def\ha{H$\alpha$}

\def\cxou{CXOU J085201.4-461753}
\def\OIII{[O\,{\sc iii}]}
\def\HII{H\,{\sc ii}}
\def\hi{\rm H\,{\sc i}}
\def\mjb{mJy beam$^{-1}$}

\def\farcs{\hbox{$.\!\!^{\prime\prime}$}}

\def\deg{\hbox{$^\circ$} }
\def\fdeg{\hbox{$.\!\!{}^\circ$}}

\def\h{^{\rm h}}

\def\m{^{\rm m}}

\def\s{^{\rm s}}

\begin{document}
\title {\textit The interior of the SNR RX J0852.0-4622\\
 (Vela Jr) at radio wavelengths}

\author {\textsc {E. M. Reynoso}\inst{1,2} 
\thanks{Member of the Carrera del Investigador Cient\'\i fico, CONICET, Argentina.}
           \and \textsc{G. Dubner}\inst{1} $^*$
         \and \textsc{E. Giacani}\inst{1} $^*$
           \and \textsc{S. Johnston}\inst{3,2} 
           \and \textsc {A. J. Green}\inst{2} }

\institute {Instituto de Astronom\'{\i}a y  F\'{\i}sica del Espacio (IAFE),
CC 67, Suc. 28, 1428 Buenos Aires, Argentina\\
             \email{ereynoso@iafe.uba.ar} 
\and School of Physics, University of Sydney, NSW 2006, Australia
\and Australia Telescope National Facility, Commonwealth Scientific and 
Industrial Research Organisation, Post Office Box 76, Epping, NSW 1710, 
Australia\\}
\offprints{E. Reynoso}

   \date{Received September 22, 2005; Accepted November 23, 2005}

\abstract{}{We observed the center of the supernova remnant Vela Jr in radio
continuum in order to search for a counterpart to the compact central X-ray 
source \cxou , possibly a neutron star candidate which could be the remnant of 
the supernova explosion.} 
{Observations were made with the Australia Telescope Compact Array at 13 
and 20 cm. Spectral indices were obtained using flux density correlations
of the data which were spatially filtered to have the same u-v coverage. A
multiwavelength search for counterparts to the compact central X-ray source
was made.}
{We compiled a new catalogue of 31 small diameter radio sources, including the 
previously known source PMN J0853-4620, listing the integrated flux densities 
at 20 cm and, for half of the sources, the flux densities at 13 cm with the 
corresponding spectral indices. All sources are unresolved at the present 
angular resolution except for Source 18, which is clearly elongated and lies 
strikingly close to \cxou . Our observations show no evidence for the 
existence of a pulsar wind driven nebula associated with the point X-ray
source. Furthermore, Source 18 has a thermal spectrum with index $\alpha =
+0.8 \pm 0.4$ ($S \propto \nu^{\alpha}$), and appears to be the counterpart
of the optical source Wray 16-30. In spite of the absence of \OIII \ emission 
lines as reported in the literature, we find that this object could be 
explained as a low emission planetary nebula belonging to the ``butterfly''
morphological class.}
{We conclude that if the radio source 18 is actually a planetary nebula, then
\cxou \ is more likely to be related to it rather than to Vela Jr.}


\titlerunning{Interior of Vela Jr at radio wavelengths}
\authorrunning{\textsc{E. Reynoso et al.}}

\maketitle

\section{Introduction}

Rotation powered pulsars can lose their energy in the form of a wind of 
electrons and positrons, creating a synchrotron emitting nebula or {\it 
plerion}, generally called a ``pulsar wind nebula'' (PWN). In the 
last few years, new data with unprecedented spatial resolution have 
revealed that PWNe are highly structured objects which can fall into a 
variety of morphological classes depending on the properties and 
evolutionary state of the pulsar and its surroundings \citep{bg04}. The 
morphology of a PWN depends on how the wind particles are confined. 
Therefore, the study of PWNe are of particular interest since their 
properties can provide information on both, the pulsar's wind and the 
surrounding interstellar medium.

In some cases, neutron stars (NS) are not detected as ordinary
radio pulsars. Several X-ray sources with no radio counterpart and very 
high X-ray to optical flux ratios have been detected projected on the
interior of supernova remnants (SNR). These sources, generally called 
``compact central objects'' \citep[CCO; see][]{pskg}, are believed to
be NS with peculiar characteristics \citep[e.g.][]
{frail,gpz}, or simply normal pulsars with unfavorable beaming
\citep{bj}. In recent papers, we have reported the results of an {\hi} \ 
and radio continuum survey towards candidate NSs to search for their 
imprints in the interstellar medium and confirm or reject their physical 
association with the host SNR \citep{iafeusyd1, iafeusyd2, geminga}. In 
this paper we analyze another CCO-SNR possible association: the X-ray 
source \cxou \  and the SNR Vela Jr. 

Vela Jr (G266.1--1.2, RX J0852.0--4622) is a shell-like SNR, about 2\deg 
in diameter, first discovered with ROSAT \citep{asch}.
The name ``Vela Jr'' is due to its position at the south-east corner of 
the Vela SNR, and to a possible age of $\sim 700$ yr based on the detection 
of $\gamma$-ray emission from the radioactive isotope $^{44}$Ti 
\citep{iyetal}. However, subsequent observations show that this source is 
likely to be older and ten times more distant (2 kpc) than was originally
believed \citep{shepmta}. \citet{crb} reported a radio counterpart based  
on the 2.42 GHz survey of \citet{dshj}. More recently, \citet{dg} find that
several features believed to be extensions to the radio shell are actually 
unrelated to Vela Jr. A detection with H.E.S.S. of TeV $\gamma$-ray extended 
emission, whose spatial distribution correlates with the X-ray observations,
has recently been reported \citep{hess}.

Several authors noted the presence of an X-ray point-like source close to 
the SNR center: \citet{asch99} based on the ROSAT All Sky Survey, 
\citet{pskg} using the Chandra Spectrometer, \citet{shepmta} based on ASCA 
data, and \citet{mere} with BeppoSAX observations. This point-like source is 
named slightly differently according to the different authors and has
independent entries listed in SIMBAD. Most frequently, it is referred to by
its Chandra identification \cxou. Due to its location, this X-ray source is
believed to be the NS remaining after the supernova explosion.
%and is classified as a CCO.  :``compact central object'' \citep[CCO; see][]{pskg}. 
This identification of \cxou \ as a NS appears to be supported
by the very high X-ray-to-optical flux ratio \citep{pskg}. 
Although the distance to \cxou \ is poorly known, if it is located at 2 kpc
the emission region and luminosity are similar to the CCO detected in Cas A. 
\citet{pelliz} found a faint \ha \ nebula at a position compatible with 
\cxou, which could be explained as a bow shock driven by a NS. 
However, \citet{iyetal2} claim that the very high ejecta velocity 
derived for Ti implies that the SN was of type Ia, in which case no 
compact remnant is expected. The nature of \cxou \ thus remains uncertain.

\citet{iafeusyd3} observed a $\sim 30^\prime$ diameter field towards
\cxou \ with the Australia Telescope Compact Array\footnote{The ATCA is part 
of the Australia Telescope, which is funded by the Commonwealth of Australia 
for operation as a National Facility managed by CSIRO} (ATCA; Frater, Brooks 
\& Whiteoak 1992) in the $\lambda$21 cm line and at 20 cm in the radio 
continuum. The observations revealed a compact continuum source and an 
elongated region ($30^\prime \times 14^\prime$) of enhanced emission at 
the position of the X-ray source. If confirmed, this could be a PWN and would 
be good evidence that the compact X-ray source is a neutron star. However, 
this enhancement is only marginally above the noise level of the image (less 
than 3$\sigma$ rms), making the detection of the nebula doubtful. In this 
paper, we analyze new observations of the central region of Vela Jr performed 
at 13 and 20 cm, which confirm the detection of the compact source. A new 
interpretation of the source origin is presented.

\medskip

\section{Observations and data reduction}

The new observations were obtained with the ATCA, the 22-m six antenna array
located in Narrabri, NSW, Australia. To optimize the $u-v$ coverage, the 
observations were carried out in 2004 during two sessions of 12 hours each, 
on February 19 in the 750A array configuration and on April 9 in the EW367 
configuration. In these two configurations, the antennas are aligned in the 
EW direction, and the baseline lengths given by the five movable antennas 
vary from 30 to 735 m. The shortest baseline and the weighting scheme 
used set an upper limit for the size of well imaged structures, estimated to 
be $\sim 35^\prime$ at 20 cm and $\sim 24^\prime$ at 13 cm, although mosaicing 
helps to partially recover short spacings and thus increase these limits. The 
inclusion of the sixth antenna, fixed in location 3 km away from the 
remainder of the array, increases the $u-v$ coverage by the addition of 
baselines from $\sim 3000$ to 4440 m. The observations were made in continuum 
mode, which allows the simultaneous measurement of two frequency bands, 
centred on 1384 MHz and 2368 MHz, each with a correlator bandwidth of 128 MHz.

The dimensions of the region of enhanced emission found in \citet{iafeusyd3}, 
required that the observations be made in mosaicing mode with six pointings 
on a hexagonal grid chosen to preserve Nyquist sampling. A seventh pointing 
was observed regularly towards the bright \HII \ region RCW 38, located $\sim 
2^\circ$ away, to assist modelling of the sidelobes from this source in the 
data reduction. The source PKS 0823--500 was observed once per hour to 
calibrate the phases, while PKS 1934--638 was used for flux density and 
bandpass calibration. Each pointing of the mosaic was reduced individually 
with the MIRIAD software package \citep{stw}, and the resulting images were 
combined as a mosaic using the MIRIAD task {\sc LINMOS}, which automatically 
applies a primary beam correction. The 13 cm images were improved by 
restricting the uvrange to less than 18 k$\lambda$. The spatial resolution of 
the finished datasets is $37\farcs 4 \times 31\farcs 7$, at position angle 
(P.A.) = $-3\fdeg 4$ for the 20 cm image, and $36\farcs 5 \times 26\farcs 3$, 
at P.A. = $-1\fdeg 7$ for the 13 cm image. The sensitivities achieved are 0.2 
and 0.1 \mjb \ respectively. 

\section {Results}

Figure 1 shows the emission at 20 cm in the central region of Vela Jr. A 
number of compact sources can be seen in the field, together with several 
filaments of faint emission. Some of the extended emission is confused by
sidelobes from RCW 38 that could not be removed.  To analyze to
what extent the extended emission in Fig. 1 is reliable, four contour 
levels have been plotted on the grayscale image at 2$\sigma$ intervals, 
starting from $2\sigma$. Most of the emission appears to be of very 
low significance, except for the filament that extends from decl.$\simeq 
-46\deg 15^\prime$ to $-46\deg 40^\prime$, at R.A.$\simeq 8\h 50\m 15\s$. 
The sidelobes around this feature are indicative that strong emission is 
present. Other features that appear genuine are those centered at about 
R.A.  $=8\h 49\m 50\s$, decl.$=-46\deg 8^\prime$ and R.A.$=8\h 51\m 40\s$, 
decl.$=-46\deg 40^\prime$ and possibly a second filament parallel to
the brightest one, at R.A.$\simeq 8\h 48\m 40\s$.  The apparent curvature
to the west of the bright filaments detected on the western half of the
image suggests a probable connection with the extended Vela SNR.

\begin{figure*}[th]
\centering
\includegraphics[width=14 cm]{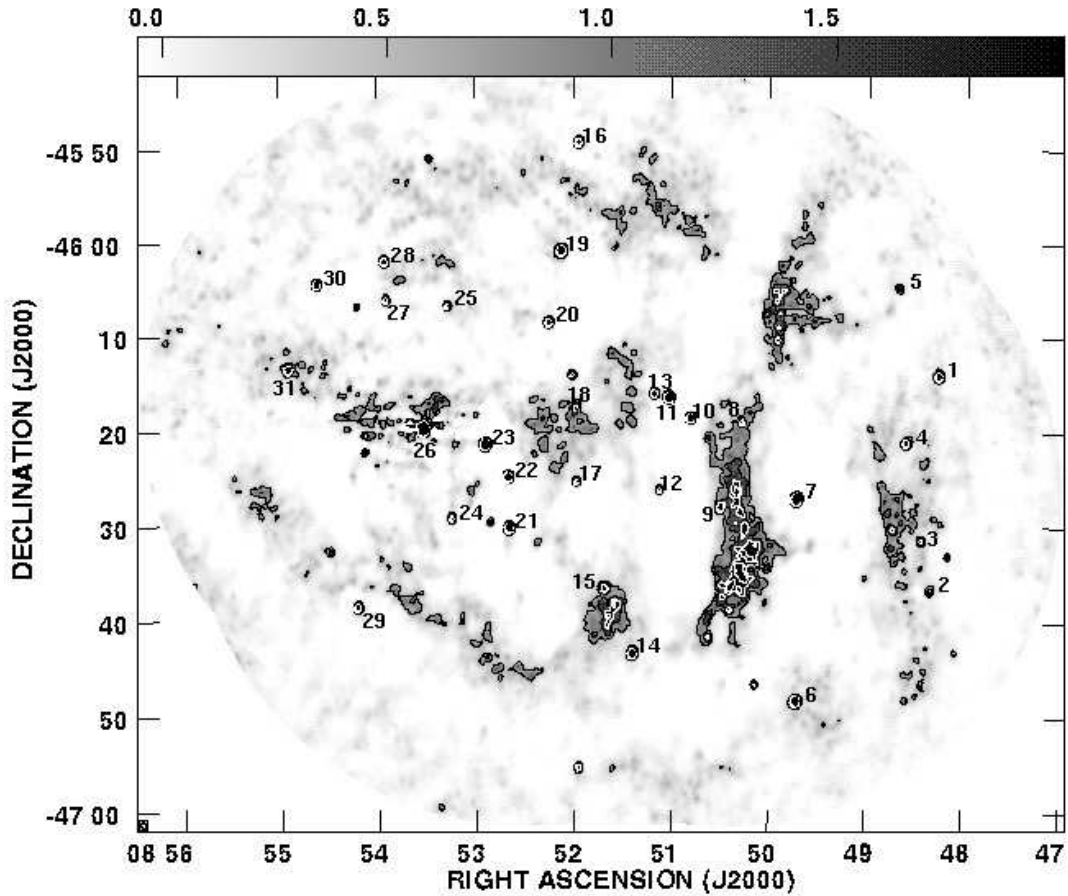}
\caption{Radio continuum image of G266.1-1.2 at 1384 MHz. The grayscale
is given on top of the image in \mjb. The contours are plotted at 2$\sigma$,
4$\sigma$, 6$\sigma$, and 8$\sigma$, where $\sigma = 0.3$ \mjb. The numbers 
correspond to the compact sources listed in Table~\ref{table1}. The beam is 
plotted at the bottom left corner.}
\end{figure*}

At 13  cm, the extended emission, if any, cannot be disentangled from
the background noise level, but half of the compact sources still appear. 
The small source reported in \citet{iafeusyd3} at a position compatible
with \cxou \ appears at both frequencies as a small, faint, elliptical
feature. In contrast with other similar features in the field, this source
is resolved. After deconvolution, it has a size of about $130\pp \times 
50\pp$ at a position angle (measured from N to E) of 170\deg at 20 cm, and 
$50\pp \times 15\pp$ at a position angle of 150\deg at 13 cm. Figure 2 shows 
a close up of a region around this source at 20 cm. The symbols correspond 
to the sources listed in Section 3.2.

\begin{figure}[th]
\centering
\includegraphics[width=8 cm]{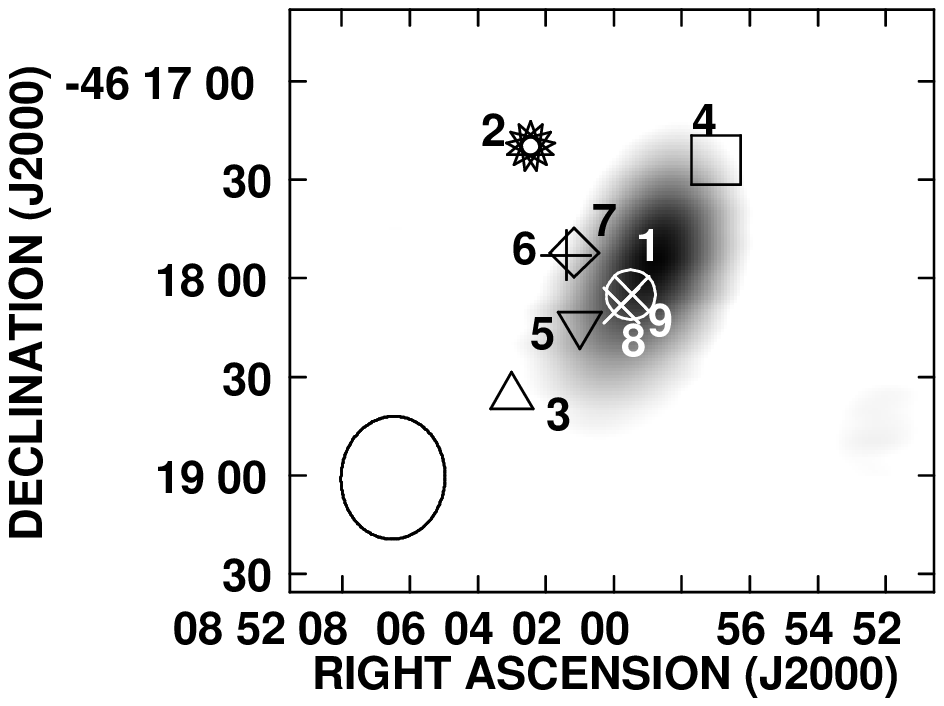}  
\caption{Close up of a $\sim 4^\prime \times 4^\prime$ region around
Source 18 (Table~\ref{table1}) at 1384 MHz. The beam is plotted at the bottom
left corner. The symbols are as follows: circle, Wray 16-30; star, HD 
76060; triangle, J085203-461836; square, AX J0851.9-4617; inverted
triangle, J085201-461814; plus sign, \cxou; diamond, EQ J0852-4617;
and crosses, MSX G266.2348-01.2004 and IRAS 0852-4606. The numbers
correspond to the sources listed in Table~\ref{table2}.} 
\end{figure}

\subsection{Small-diameter sources in the field}

To investigate the nature of the compact sources appearing in the field, we 
estimated their positions and fluxes at both frequencies and calculated the 
spectral index between 13 and 20 cm where possible. The results are summarized 
in Table~\ref{table1}. The coordinates were obtained from a Gaussian fit to 
the flux density peak. When necessary, fluxes were corrected for contamination 
from background extended emission; this effect is more significant at 20 cm.
The spectral indices were derived for each source from the comparison of 
the two images using the flux-flux plotting technique \citep{costain,turtle}. 
This method compares the flux densities at two frequencies within a selected 
region of the sky for which the data have been spatially filtered for the same 
$u-v$ coverage. A linear fit to the plot gives a measure of the spectral index. 
This method is useful because it is unaffected by absolute calibration and 
offset variations between the images. 

\begin{table}
\caption{Small-diamter sources in the field}
\label{table1}      
\centering          
\begin{tabular}{cccccc}
%  & \multicolumn{3}{c}{~~~~~~~~~~~~~~~~Table 1. Small-diamter sources in
%the field}\\
\hline
\hline
& RA (J2000) & Dec. (J2000) &S$_{\rm 13cm}$&S$_{\rm 20cm}$&~Spectral\\ 
%& ~~\hh~~ \mm ~~ \ss & ~~\d ~~~\m ~~~\s & (mJy)& (mJy)& index \\ 
& ~~h~~ m ~~ s & ~~$^\circ$ ~~~$^\prime$ ~~~$^{\prime\prime}$&(mJy)&(mJy)&index\\ 
\hline

1 &08 48 16.2 & -46 14 11& &8.4& \\
2 &08 48 20.9 & -46 36 57& &2.0& \\
3 &08 48 26.5 & -46 31 35& &2.8& \\
4 &08 48 36.0 & -46 21 22& &6.7& \\
5 &08 48 40.4 & -46 04 58& &2.8& \\
6 &08 49 42.7 & -46 48 36& &20& \\
7 &08 49 42.8 & -46 27 19& 28&49&$-$0.98$\pm$0.04 \\
8 &08 50 16.8 & -46 19 22& 1.8 &2.1&$-$0.35$\pm$0.10 \\
9 &08 50 29.9 & -46 28 17& 4.2 &5.0&$-$0.6$\pm$0.2 \\
10 &08 50 48.2 & -46 18 54& 8.3 &15&$-$1.2$\pm$0.1 \\
11&08 51 01.4 & -46 16 43& 33 &31&0.21$\pm$0.02 \\
12&08 51 07.1 & -46 26 30& 3.5 &3.2&0.12$\pm$0.13 \\
13&08 51 10.4 & -46 16 17& 4.9 &8.0&$-$0.7$\pm$0.2 \\
14&08 51 23.6 & -46 43 43& 2.3 &16& \\
15&08 51 41.0 & -46 36 51& &4.6& \\
16&08 51 56.8 & -45 49 37& &7.1& \\
17&08 51 58.1 & -46 25 35& 3.6 &3.1&0.23$\pm$0.07 \\
18&08 51 58.9 & -46 17 52&3.4&2.4&0.8$\pm$0.4 \\
19&08 52 07.3 & -46 01 16& 2.7 &19& \\
20&08 52 15.1 & -46 08 48& 9.7 &8.1&0.35$\pm$0.01 \\
21&08 52 39.5 & -46 30 30&10 &18&$-$0.95$\pm$0.07 \\
22&08 52 40.1 & -46 25 07& 12 &17&$-$0.54$\pm$0.05 \\
23&08 52 54.1 & -46 21 42& 23 &39&$-$0.92$\pm$0.04 \\
24&08 53 14.6 & -46 29 36& 2.1 &3.9&$-$1.1$\pm$0.3 \\
25&08 53 17.1 & -46 06 59&  &2.8& \\
26&08 53 31.9 & -46 19 59&192 &226&$-$0.34$\pm$0.02 \\
27&08 53 54.7 & -46 06 21& &3.8& \\
28&08 53 55.6 & -46 02 15& &7.4& \\
29&08 54 12.6 & -46 38 45& &7.1& \\
30&08 54 36.9 & -46 04 38& &13& \\
31&08 54 54.6 & -46 13 38& &10& \\
\hline
\end{tabular}
\end{table}

To apply the flux-flux plots method, the image at 13 cm was convolved to the 
20 cm beam size. To allow for poor correlation coefficients, the regression 
was done twice, by plotting data at 13 cm vs. 20 cm and vice versa. The 
spectral indices shown are the average of both estimates. The correlation 
coefficient was nearly one for almost all the sources. The quoted errors 
include a contribution from the intrinsic flux density errors plus the 
statistical error from the linear regression. Estimates were made only for 
sources with a flux density $>$ 1 \mjb \ that were at least 3$\sigma$ above 
the image noise level. As an example, flux-flux plots for two representative 
sources (18 and 26) are shown in Figure 3.

\begin{figure}[th]
\centering
\includegraphics[width=8 cm]{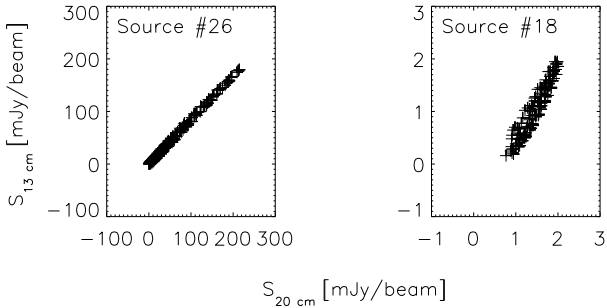}
\caption{Flux-flux plots for sources 18 (right) and 26 (left), where the
flux density at 13 cm is plotted against the flux density at 20 cm. Units
are \mjb.}
\end{figure}

For sources 14 and 19, we do not list the spectral indices because they lie 
very close to the boundary of the calibrated image at 13 cm, and their flux
densities are incompletely recovered, which produces unrealistic values for
the spectral index.

After searching the available published radio catalogs, we find that  the
only compact source  previously identified in this field is number 26 in
Table~\ref{table1}, listed in the Parkes-MIT-NRAO Survey \citep{wright} as 
PMN J0853-4620, with a flux density of 151 mJy at 4.85 GHz. The spectral
index derived for this source, including the PMN data, yields exactly the 
same result as using the flux-flux plot, giving us confidence in the
reliability of the technique.

The region observed for this paper has been partially imaged in the
Molonglo Synthesis Telescope Galactic Plane Survey at 843 MHz with sub
arcmin resolution (Green et al. 1999). The sources and extended structures
are consistent with the present data. Two images have also been published by
\citet{stupar} at 13 and 20 cm, which are less sensitive but in broad agreement.

About half of the sources for which spectral indices have been
computed appear to be extragalactic objects ($\alpha$ more 
negative than $-$0.9).  The remaining sources have spectral indices 
comprised between $\sim -0.7$ and +0.8. Source 18 is the only resolved 
source and is the target of our interest because of its proximity to 
\cxou. It has the highest positive spectral index in the field and we now
discuss the possible origin of the emission associated with this source. 

\subsection {Multi-wavelength data possibly related to the central source}

We have searched in the SIMBAD Astronomical Database for sources of
all categories located at or close to the position of the  radio nebula, 
Source 18. At least five X-ray sources, four optical objects and one radio 
source, are listed within a 1$^\prime$ radius circle centered at the position 
08\hh 52\mm 00\ss.0, $-46\deg 17^\prime 54\pp$. However, there are 
inconsistencies and possible duplications in the classifications from the 
published data. In an attempt to clarify the nomenclature we have listed in 
Table~\ref{table2} the catalogued objects with their original names and 
positions as published by the respective authors, following in more or less 
historical order. We have excluded those sources labeled in SIMBAD as SNR RXJ 
0852.0-4622, GRO J0852-4642, and [CRB99] A, which correspond to the X-ray, 
$\gamma$-ray, and radio manifestations of the SNR Vela Jr. A summary of these 
findings follows.

(1) Wray 16-30 was originally listed by \citet{wray} as a Be 
star, with coordinates accurate to about $\pm 15\pp$. This source
is also listed in the SIMBAD database as Ve 2-27, an object originally
classified as a planetary nebula (PN) with star-like appearance by 
\citet{velghe}, and later as a  peculiar Be star with infrared excess 
\citep{ayg}. In the Strasbourg-ESO Catalogue of Planetary Nebulae 
\citep{acker}, it is classified as a  peculiar emission-line source 
(see \citeauthor{landaetal} \citeyear{landaetal} for a summary of the 
successive classifications of this emission-line  object). 
\citet{landaetal} find that Ve 2-27 (or Wray 16-30) is a peculiar 
emission-line star with a spectrum remarkably similar to that of the 
Luminous Blue Variable object $\eta$ Carinae, but with an unclear
evolutionary status. However, according to the spectral study carried out
by \citet{landaetal},  it is still possible to explain the observed spectrum
as originating from a proto-planetary nebula.

(2) HD 76060 is catalogued in the Hipparcos Catalogue as a bright
(B=7$^m$.8) B8IV/V star.

\begin{table*}
\caption{Multi-wavelength detections near Source 18}
\label{table2}
\begin{tabular}{lllllc}
%  & \multicolumn{3}{c}{~~~~~~~~~~~~~~~~Table2. Multi-wavelength detections
%near Source 18}\\
\hline
\hline
Number & Identifier & RA (J2000) & Dec. (J2000) & Other names & References\\ 
& & ~h~ m ~ s & ~$^\circ$ ~$^\prime$ ~$^{\prime\prime}$ & & \\ 
\hline

1 & Wray 16-30& 08 51 59.5 & $-$46 18 05 & ESO 260-PN11, Hen 2-14,& 1,2,3 \\
& & & & Ve 2-27, PK 266-01 1, SS73 11 & \\
2 & HD 76060 & 08 52 02.45 & $-$46 17 19.8 & & 4\\ 
3 & [AIS99] J085203-461836 & 08 52 03 & $-$46 18 36 & & 5\\
4 & AX J0851.9-4617 & 08 51 57 & $-$46 17.4 & & 6\\
5 & [M2001b] J085201-461814 & 08 52 01 & $-$46 18 14 & & 7\\
6 & CXOU J085201.4-461753 & 08 52 01.4& $-$46 17 53 & & 8\\
7 & EQ J0852-4617 & 08 52 01.17 & $-$46 17 52.3 & & 8\\
8 & MSX G266.2348-01.2004 & 08 51 59.8 & $-$46 18 8.28 & & 9\\
9 & IRAS 08502-4606 & 08 51 59.5 & $-$46 18 4.68 & & 10\\
\hline
\end{tabular}

{{\bf References:} (1) \citeauthor{velghe} \citeyear{velghe}; (2) 
\citeauthor{landaetal} \citeyear{landaetal} and references therein; 
(3) \citeauthor{wray} \citeyear{wray}; (4) Hipparcos Catalogue, 
\citeauthor{perryman} \citeyear{perryman}; (5) \citeauthor{asch99} 
\citeyear{asch99}; (6) \citeauthor{shepmta} \citeyear{shepmta}; (7) 
\citeauthor{mere} \citeyear{mere}; (8) \citeauthor{pskg} 
\citeyear{pskg}; (9) MSX6C Infrared Point Source Catalog \citeauthor{egan}
\citeyear{egan}; (10) IRAS Point Source 
Catalogue }

\end{table*}

(3) [AIS99] J0825203-461836 refers to a point-like source 
discovered in the ROSAT all sky survey data by \citet{asch99}. It is about 
$3^\prime.4$ offset from the center of Vela Jr. A second point-like X-ray 
source is reported near $08\h 51\m 58\s$, $-46\deg 21^\prime 33\pp$. The 
authors discuss the possiblity that one of these sources could be a neutron
star. However, since no spectra were available at that time, 
the analysis was incomplete.

(4) AX J0851.9-4617 was detected with ASCA as an unresolved X-ray source
surrounded by diffuse emission and located  near the center of SNR RXJ
0852.0-4622 \citep{shepmta}. According to the authors, the properties of
the source are not well determined but appear to be consistent with a
neutron star surrounded by a synchrotron nebula, although an association
with one of two stars located within the positional error circle is also
possible. 

(5) [M2001b] J085201-461814 is the X-ray source that
\citet{mere} reported to be at the center of the SNR RX J0852.0-4622. 
It should be noted that at energies E$\geq$ 5 keV a  new source was
detected, indentified as SAX J0852.0-4617, 
which has a harder X-ray spectrum and higher X-ray-to-optical flux 
ratio than was measured for the original neutron star candidate AX 
J0851.9-4617. This new source is a more likely candidate for the central 
neutron star.

(6) \cxou \ is a point X-ray source discovered with the 
Advanced CCD Imaging Spectrometer detector aboard {\it Chandra}. 
The source lies about $4^\prime$ north of the center of the SNR RXJ
0852.0-4622. \citet{pskg} proposed that it is the compact remnant 
of the SN explosion.

(7) EQ J0852-4617 is a faint (B$\sim 19^m$) star reported 
by \citet{pskg}. It was discovered in a search for an optical counterpart 
to the point-like X-ray source \cxou , although it has now been rejected as
such. It is offset by about $2\pp.4$ from the nominal X-ray position.

We have also looked for an infrared counterpart in images from the 
Two Micron All Sky Survey (2MASS), the MSX6C Infrared Point Source 
Catalog \citep{egan}, and the IRAS Sky Survey Atlas in the bands 1.25 
(J), 1.65 (H), 2.17 (K), 8.28 (A), 12.13 (C), 14.65 (D), 21.3 (E), 12, 
25, 60, and 100 $\mu$m. In all three surveys, an unresolved source is found 
at the position of Source 18. The infrared sources included in the MSX and 
IRAS catalogues are listed in Table~\ref{table2} as numbers 8 and 9 and are 
plotted as crosses in Fig. 2.

\medskip

\section {Discussion}

\subsection{The nature of Source 18}

Source 18 is a resolved source in the
field with the appearance of a small, elliptical nebula. In spite of 
its proximity to \cxou, their relationship is not clear. The area 
subtended by Source 18 at 20 cm, after beam deconvolution, marginally 
encloses \cxou, which lies $26\pp$ away from the radio emission peak. 
At 13 cm, \cxou \ is well separated from the radio emission. 
The spectral index derived for Source 18 is strongly thermal, which makes
it unlikely to be a PWN created by \cxou. The existence of infrared
counterparts reinforces the interpretation of Source 18 as a thermal
source.

\citet{wb74} showed that an ionized, uniform, spherically symmetric
mass loss flow leads to a spectrum with an index equal to 0.6 in the 
radio and infrared regimes. For example, the radio star P Cygni has a 
spectral index of 0.65. Within the uncertainties, the spectral index 
derived for Source 18 is the same. Similar values have also been found for
other radio stars.  
Therefore, Source 18 is probably an object with significant mass loss. 
The [12]-[25] and [25]-[60] color indices of this object, as derived from 
the fluxes listed for IRAS 08502-4606, are 0.12 and 0.27 respectively. 
These values place Source 18 in the region VIb of the \citet{vdvh88}
two-color scheme, which corresponds to objects with separate hot and
cold dust components indicating that the mass loss process has been 
significantly interrupted. \citet{vdvh88} conclude that objects in
the VIb region are variable stars of which a small number have hot 
oxygen-rich material distributed separately from their cold dust
component. 

In optical images, Source 18 lies at the same location as Wray 16-30.
Figure 4 shows the red ESO/R/MAMA image \citep[downloaded from the 
Aladin Sky Atlas;][]{bonnareletal00}, centered on Vela Jr, with radio 
continuum contours at 13 cm corresponding to Source 18 superimposed. The 
black star (pointed by an arrow) indicates the location of \cxou, while  
the white crosses correspond to the infrared peaks (according to the IRAS
and MSX catalogues). In optical emission, Wray 16-30 appears complex,  
with two bright lobes symmetrically located on either
side of a fainter linear filament with a NE-SW orientation. The 
whole structure is particularly bright in the \ha \ line \citep[cf.]
[]{pelliz}. The IR peaks lie on the brightest of the two lobes. 
The two lobes are enclosed within Source 18, with the 
jet-like axis of the optical nebula oriented orthogonally to the 
major axis of the radio nebula. The optical axis runs
across the radio emission peak. We strongly believe that Source 
18 is the radio counterpart of Wray 16-30. Higher
resolution images would be very valuable to investigate the
morphology at radio wavelengths in more detail.

\begin{figure}[th]
\centering
\includegraphics[width=8 cm]{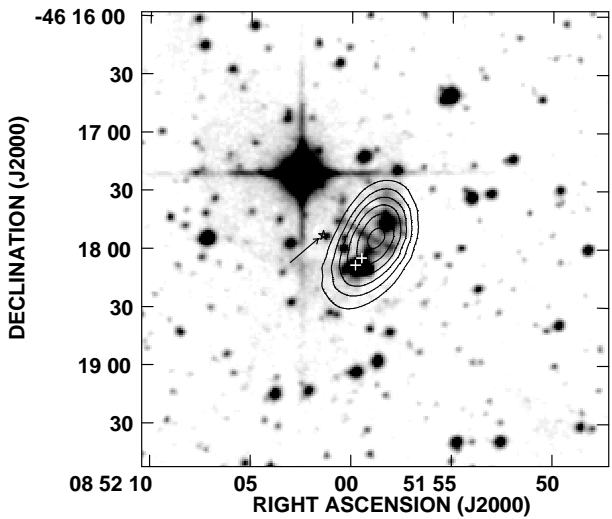}
\caption{Red ESO/R/MAMA image of the center of Vela Jr. The black 
star (pointed by an arrow) shows the position of \cxou, while the two 
white crosses indicate the IR source IRAS 0852-4606 as given in 
the MSX and IRAS catalogues. A few radio contours at 13 cm are included, 
which correspond to Source 18 (see Table~\ref{table1}).}
\end{figure}

The optical morphology of Wray 16-30 resembles that of ``butterfly''
planetary nebulae \citep[e.g.][and references therein]{icke99}.
However, there is no agreement in the literature on the classification of 
Wray 16-30. As summarized in Section 3, references interpreting it as a 
peculiar PN or rejecting it as such are found almost equally. The 
problem arises from the absence of \OIII \ emission lines. For example, 
\citet{styac} strictly reject as PNe those objects for which the ratio 
$I$\OIII$/I(\rm H \beta) =0$. Conversely, \citet{kohoutek} classifies 
them as PNe of excitation class 0 (corresponding to very-low-excitation 
objects, \citeauthor{ss} \citeyear{ss}), provided additional indicative 
evidence is found. In Source 18, such evidence could be the elliptical 
morphology, the non-stellar size, and the thermal radio continuum 
spectrum, typical of the vast majority of PNe (although non-thermal 
spectra are seen in some cases; e.g. \citeauthor{isaacman}
\citeyear{isaacman}). The low excitation class contains objects which 
are evolving into typical PNe \citep{kohoutek}. 

If Source 18 (or Wray 16-30) is indeed a PN, a physical association
with \cxou \ still cannot be ruled out. X-ray emission has been 
detected towards several ``butterfly'' PNe \citep{feibelman,kastner}.
\citet{wgmck} found many X-ray sources strikingly close to PNe in 
M31. Although the X-rays are not coming from the PN themselves 
(they lie several arcseconds away), the authors conclude that there is 
a physical connection between them, and suggest that these associations 
may involve a PN and a low-mass X-ray binary occupying the same 
undetected star cluster. If this is the case, then \cxou \ would not 
be the neutron star associated with Vela Jr, in accordance with the recent
suggestion by \citet{iyetal2}. 

To further investigate a possible link between the PN and the X-ray source,
we check to see if the distance to both sources are similar based on their
extinction values. \citet{landaetal} estimated the colour excess of Wray 
16-30 to be $E(B-V)=0.93 \pm 0.45$ mag. The reddening to \cxou \ can be 
derived from the \hi \ column density which, based on XMM-Newton observations,
is $N_H=(3.7-4.3)\times 10^{21}$ \c2 \ \citep{basch}. Using the ratio 
$N_H/A_J=(5.6\pm 0.4)\times 10^{21}$ \c2 \ mag$^{-1}$ \citep{vuong} and 
the standard values $A_V/A_J = 3.31$ and $R=A_V/E(B-V)=3.1$ for diffuse 
dust \citep[e.g.][]{mathis}, the colour excess of \cxou \ turns out to be 
$0.76\pm 0.12$ mag. To apply the reddening-distance model of \citet{chen}, 
we first estimated the total reddening  in the direction of \cxou \ produced 
by the Galactic Plane along the line of sight to be 2.15 mag \citep{ebvmaps}. 
The derived distances to Wray 16-30 and \cxou \ are thus $3.0\pm ^{2.2}_{1.5}$ 
kpc and $2.4 \pm 0.4$ kpc respectively. Therefore, considering the 
uncertainties, it is possible that both sources lie at the same distance. 

\medskip
\subsection{Is there a PWN associated with \cxou?}

An alternative way to test if \cxou \ is a neutron star is by 
searching for the radio counterpart corresponding to a PWN. Although 
no conclusions can be drawn from a negative result, a positive PWN 
detection is a strong indication that the powering source must
be a neutron star.

Two radio features were suggested by \citet{iafeusyd3} as a possible PWN
blown by \cxou. One of them was Source 18. From the discussion above, it is
clear that this interpretation is unlikely. The other candidate was a
$30^\prime$ long nebulosity. The new observations presented here reveal
that this feature is, in fact, an artifact due to poor $u-v$ \ coverage in 
the interferometric radio data.

In Fig. 1, \cxou \ lies on a local enhancement of diffuse emission. 
This region is not larger than a few arcmin in diameter and its
emission never exceeds the 3$\sigma$ level, making it very difficult 
to establish its significance unambiguously. The integrated flux density of
this feature using a boundary set rather subjectively is 76 mJy at 20
cm. If this feature were the PWN driven by \cxou, then  
it would be characterized by a flat non-thermal spectrum \citep[see 
for example][]{elsapwn}. Assuming a spectral index flatter than $-$0.3,
then the mean flux density at 13 cm over the same region (and with the same
spatial filtering) should be $\gtrsim 4$ \mjb. However, the measured flux 
density at this wavelength is only $\sim 1$ \mjb. Therefore, either 
the central emission at 20 cm is not a genuine source or it has a spectrum
steeper than the expected model. In either case, no PWN is indicated.

\section{Conclusions}

We have made a high resolution, high sensitivity image of the center of the
SNR Vela Jr at 13 and 20 cm. The aim of the project was to detect a radio
counterpart to the X-ray source \cxou, a neutron star candidate for the
compact remnant of the SN explosion. There is no evidence for the existence
of a PWN associated with \cxou. The closest radio feature detected is a
compact source of thermal origin, named Source 18 in this paper, which
appears to be the counterpart of the optical source Wray 16-30 detected
also at infrared wavelengths as IRAS 08502-4606. In spite of the lack of
\OIII \ emission lines, we believe there is good evidence that Wray
16-30 is a PN, probably belonging to the morphological class known 
as ``butterfly'' PNe. If so, it is also possible that \cxou \ is
physically associated with it. In that case, \cxou \ would then not be 
the CCO of SNR Vela Jr. This interpretation is strengthened by the radio
frequency results presented here, which do not indicate an association
between \cxou \ and Vela Jr. 

The angular resolution of the present observations is insufficient to show  
whether Source 18 has an internal structure similar to the optical 
images. Higher resolution data could be obtained with the ATCA in an
extended configuration and at a higher frequency.

A by-product from this study is the discovery of 30 new, uncatalogued
unresolved sources, which may be useful for comparison with high energy
searches. For almost half of these sources we were able to measure the flux
densities at both frequencies, 1348 and 2368 MHz, and have computed 
spectral indices using the flux-flux method. The remaining sources were too
weak at 13 cm to derive reliable results. Among those sources for which
spectral indices could be measured, six were found to be extragalactic and
five are clearly thermal, including the previously known source PMN
J0853-4620. The only one with a spectral index close to 0.6, typical of
stars with mass outflows, is Source 18.
  
\bigskip

\begin{acknowledgements}
The authors wish to thank the referee, Tyler Foster, for his very useful
comments which helped to improve the quality of the paper.
This project was partially financed by grants ANPCyT-14018 and UBACYT A055. 
During part of this work, E. M. R. was a visiting scholar at the University 
of Sydney. This research has made use of Aladin. 
\end{acknowledgements}

\label{lastpage}
\end{document}